\documentclass{ws-ijmpd}
\usepackage[super,compress]{cite}

\usepackage{amsmath}
\usepackage{comment}
\usepackage{graphicx, float}
\usepackage{dcolumn}
\usepackage{bm}
\usepackage{epstopdf}
\usepackage{doi,hyperref,url}
\begin{document}

\markboth{D. Barta \& M. Vas\'{u}th}
{Erratum: Dispersion of gravitational waves in cold spherical interstellar medium}

%
\catchline{}{}{}{}{}
%

\title{
	Erratum \\ Dispersion of gravitational waves in cold spherical interstellar medium \\[10pt] 
	\begin{small}
	\normalfont [International Journal of Modern Physics D Vol. 27, No. 4 (2018) 1850040]
	\end{small}
}

\author{D\'{a}niel Barta${}^{*}$ and M\'{a}ty\'{a}s Vas\'{u}th${}^{\dagger}$}

\address{Institute for Particle and Nuclear Physics, Wigner Research Centre for Physics,\\
	Konkoly-Thege Mikl\'{o}s \'{u}t 29--33, Budapest, 1121, Hungary\\
	${}^{*}$barta.daniel@wigner.hu\\
	${}^{\dagger}$vasuth.matyas@wigner.hu}

\maketitle

\begin{history}
\received{27 August 2020}
\end{history}

\begin{abstract}
The study published in \textit{IJMPD} \textbf{27}(4):1850040, 2018 provided a numerical result for the frequency-shift of GWs due to dispersion in interstellar medium. In order to adjust the metric functions of the originally improperly matched `background' spacetime in sections 2.1, the authors have adopted Darmois--Israel junction conditions. In section 4.1 the code used in the original paper erroneously computed the magnitude of frequency-shift for the transient event GW150914 due to a missing conversion factor. In both cases where numerical errors and potential contradictions have been identified and eliminated, adjustments were undertaken in order to maintain consistency with closely-related earlier studies.\\
\end{abstract}

Sec 2.1 set forth the `background' spacetime of a spherically symmetric giant molecular cloud (GMC) wherein the high-frequency gravitational waves (GWs), represented by metric perturbations, were embedded in the original paper\cite{barta2018}{}. In view of the importance of the study of self-gravitating isothermal gas spheres and the demand for a full analytical solution within the framework of general relativity, we introduce a new approximate analytical solution in our preceding paper \cite{barta2013}{}. The method proposed for finding an approximate analytic solution was primarily based on a transformation of the field equation into a quadratic algebraic equation in some generating function of the pressure, of the density and of the metric potential, and a set of model parameters was determined by matching the interior fluid solution to the static vacuum exterior spacetime across a hypersurface separating the corresponding Lorentzian manifolds. However, in order to obtain the parameters $A$, and $C$ that characterize the interior fluid solution (15) in the aforementioned paper, the internal manifold $\mathcal{M}^{-}$ has to matched smoothly with the static exterior spacetime $\mathcal{M}^{+}$ by locally deforming the associated ambient spacetime metrics in relation to each other\cite{Gaspar2010}{}. In agreement with the well-known Darmois--Israel (or Israel--Sen--L\'{a}nczos--Darmois) junction conditions \cite{Israel1966}{}, joining the two spacetime partitions $(\mathcal{M}^{\pm},\, g^{\pm})$ associated with the Lorentzian manifolds $\mathcal{M}^{\pm} = \mathcal{M}^{\pm} \cup \partial\mathcal{M}^{\pm}$ requires the induced metric and extrinsic curvature to be continuous across a space- or timelike boundary hypersurface $\Sigma$.\cite{Keresztes2010}{} In line with Huber's study\cite{Huber2019}{}, the manifold of the combined spacetime $\mathcal{M} = \mathcal{M}^{+} \cup \partial\mathcal{M}^{-}$ is the union of the manifolds of the individual parts, in order to $\Sigma \subset \partial\mathcal{M}^{\pm}$ be a part of the boundary of both spacetimes. Considering that the spherical clouds of isothermal gas are highly non-relativistic, the metric components of the `appropriately matched' spacetime they reside in are nearly flat, and, therefore, barely differ from those matched with the metric components of the Minkowski spacetime on the boundary. \\

In Sec 4.1, as a numerical solution to the transport equation of secondary amplitudes, the frequency shift due to the absorption of gravitational radiation by the medium has been found to reach $\hat{f} \sim -10^{-11} $ Hz, which differs from the results of Thorne\cite{Thorne1997}{} and Loeb\cite{Loeb2020}{} by seven orders of magnitude. In order to make our results comparable to those of the aforementioned earlier studies, we also calculate an upper limit on the dissipated fraction of the energy carried by GWs in a similar fashion. From a different point of view, Thorne\cite{Thorne1997}{} stated that the rate of loss of GW energy is precisely equal to the rate of heating of the viscous medium where the GW-induced shear stress is working against the viscosity. The process produces heat that increases the background energy density of the medium.

The local speed of sound in the medium is determinded as $c_{\text{s}}^{2} = dp/d\rho$ where a radial profiles of density and pressure are shown by Fig. 1 in our earlier paper\cite{barta2013}{}. The average speed of sound in the gaseous interstellar medium as cold as $T = 10$ K is $c_{\text{s}} \approx 0.3$ km/s. As a result, the one-dimensional velocity dispersion in the medium is $\sigma = \frac{1}{2}\sqrt{\langle v^{2} \rangle} = 0.1$ km/s, whereas the dynamical time in sphere with a radius of $R = 70$ pc is around $t_{\text{dyn}} \sim R/\sigma \approx 2.16 \times 10^{16}$ s. Considering that the fraction of energy dissipated from the waves by the medium is
\begin{equation}
\epsilon_{\text{diss}} < 36\left(\frac{\sigma}{c}\right)^{3}\left(\frac{1}{\omega t_{\text{dyn}}}\right),
\end{equation}
as stated by Loeb\cite{Loeb2020}{} in his eq. (12), the upper limit is $\epsilon_{\text{diss}} <1.34 \times 10^{-18}$. By the restoration of $c$ and $G$ factors which, for our convenience, were dropped throughout the original paper, one finds that reinserting a factor of $G^{-1}c^{-2}$ to the right-hand side of eq. (36) accounts for the factor of $\mathcal{O}(10^{-7})$ missing from $\hat{\omega}$ and yields the appropriate dimension $1/s$.

Finally, the authors wish to additionally provide an intuitive and physically motivated interpretation of the obtained results. Prior to Ehlers \& Prasanna studies\cite{Ehlers1987,Ehlers1996}{}, which have been the primary sources of our paper, Sacchetti \& Trevese \cite{Sacchetti1979}{} pointed out that the results of the second-order WKB analysis for the dispersion of GWs ``may well be interpreted as zero-temperature dispersion effects due to the gravitational polarization of matter''. In his above cited study, Thorne\cite{Thorne1997}{} concludes that whenever the energy of GWs passing through interstellar matter gets absorbed by the medium, chromatic dispersion occurs owing to the shear stress that damps the waves. In other words, their frequency ($\omega$) or wavelength ($\lambda$) depends on the phase velocity, defined by 
\begin{equation}
v_{\text{ph}} = \partial \omega/\partial k,
\end{equation}
and is well-approximated by $v_{\text{ph}} \approx 1 + \lambda^{2}/\mu$ and the dispersion relation is expressed by
\begin{equation}
\omega^{2} = k^{2} + 16\pi(\mu-i\omega\eta)
\end{equation}
for GWs of sinusoidal form $h^{\text{GW}} \propto e^{i(kz - \omega t)}$ propagating with a frequency $\omega$ and spatial frequency (i.e. wavenumber) $k$ along the $z$-axis\cite{Thorne1997}{}.  The term $16\pi|\mu-i\omega\eta|$ is extremely small, $\sim \mathcal{O}(10^{-33})$, compared to $\omega^{2} \cong k^{2} \cong \lambda^{2}$, where $\mu$ and $\eta$ stand for the dynamic viscosity and the coefficient of shear viscosity, respectively\footnote{The typical values of viscosity for ordinary solid bodies are extremely small, $\mu \sim 10^{-36}$ cm${}^{-1}$ and $\eta \sim 10^{-39}$ cm${}^{-2}$, let alone for much less denser GMCs\cite{Thorne1997}{}.}.

\section*{Acknowledgments}
We are most grateful to Dr. Zolt\'{a}n Keresztes and Dr. Bence Kocsis who, as LIGO members as well as opponents for the lead author's doctoral thesis defence at the E\"{o}tv\"{o}s Lor\'{a}nd University, have brought some of the inconsistencies in the original paper to our attention. We sincerely appreciate all the valuable comments and suggestions, which greatly helped us in the identification of inconsistencies and errors. 

\bibliographystyle{ws-ijmpd}
\bibliography{ws-ijmpd}

\end{document}